\documentclass{PoS}

\title{Two-pion interferometry for the granular sources in the heavy ion
collisions at RHIC and LHC energies}

\ShortTitle{Two-pion interferometry for the granular sources ...}

\author{\speaker{Wei-Ning Zhang}\\
School of Physics and Optoelectronic Technology,
Dalian University of Technology, Dalian, Liaoning 116024, China\\
Department of Physics, Harbin Institute of Technology, Harbin,
Heilongjiang 150006, China\\
E-mail: \email{wnzhang@dlut.edu.cn}}

\abstract{We investigate the two-pion interferometry in ultrarelativistic
heavy ion collisions in the granular source model of quark-gluon
plasma droplets.  The pion transverse momentum spectra and HBT radii
of the granular sources agree well with the experimental data of the
most central Au-Au collisions at $\sqrt{s_{NN}}=200$ GeV at RHIC and
Pb-Pb collisions at $\sqrt{s_{NN}}=2.76$ TeV at LHC.  In the granular
source model the larger initial system breakup time for the LHC
collisions as compared to the RHIC collisions may lead to the larger
HBT radii $R_{\rm out}$, $R_{\rm side}$, and $R_{\rm long}$.  However,
the large droplet transverse expansion and limited average relative
emitting time of particles in the granular source lead to the ratio
of the transverse HBT radii $R_{\rm out}/R_{\rm side} \sim 1$.}

\FullConference{The Seventh Workshop on Particle Correlations and Femtoscopy\\
		 September 20 - 24 2011\\
		 University of Tokyo, Japan}

\begin{document}

\section{Introduction}
Hanbury-Brown-Twiss (HBT) interferometry is a useful tool to probe
the space-time geometry of the particle-emitting sources in high
energy heavy ion collisions \cite{CYW94,UAW99,RMW00,MAL05}.  The
experimental results of the HBT measurements for the Au-Au
collisions at the high energies of the Relativistic Heavy Ion
Collider (RHIC) indicate that it is hard to describe the source
space-time dynamics by a simple evolution model
\cite{STA01a,PHE02a,PHE04a,STA05a}.  HBT interferometry data provide
strong constraints for the models of source space-time dynamics.
Recently, the HBT measurement for the $\sqrt{s_{NN}}=2.76$ TeV Pb-Pb
most central collisions at the Large Hadron Collider (LHC) has been
performed \cite{ALI11}.  A consistent explanation to the HBT data of
the LHC and RHIC experiments is required naturally for the source
models, which will be helpful to understand the initial condition,
source evolution, and particle freeze-out in ultrarelativistic heavy
ion collisions.

In Refs. \cite{WNZ06,WNZ09}, the granular source model of
quark-gluon plasma (QGP) droplets \cite{WNZ04} is developed to
explain the RHIC HBT data \cite{PHE04a,STA05a}.  In this work we
investigate the two-pion HBT interferometry in ultrarelativistic
heavy ion collisions in the granular source model of QGP droplets.
Our results indicate that the granular source for the LHC Pb-Pb
collisions may have a larger initial system breakup time as
compared to the RHIC Au-Au collisions.  The granular source model
consistently reproduces the pion transverse momentum spectra and HBT
radii in the most central collisions of the RHIC
\cite{PHE04,STA04,STA05a} and LHC \cite{ALI11a,ALI11} experiments.

\section{Granular source model in ultrarelativistic heavy ion collisions}
In ultrarelativistic heavy ion collisions, the system at central
rapidity may reach a local equilibrium at a very short time
$\tau_0$, and may then expand rapidly along the beam direction
($z$-axis). Because of the initial fluctuation, the initial
local-equilibrium system is not spatially uniform \cite{HJD02OSJ04}.
It may form many tubes along the beam direction during the fast
longitudinal expansion, and finally fragment into many droplets
(see Fig.1 of Ref. \cite{WNZ06}) due to the ``sausage" instability
and surface tension \cite{WNZ06}.  On the other hand, the rapidly
increased bulk viscosity in the QGP near the phase transition may
also leads to the breakup of the system \cite{GTO08}.

We assume that the system fragments and forms a granular source of
many QGP droplets at a time $t_0 (>\tau_0)$.  On the basis of the
Bjorken hypothesis \cite{JDB83}, the longitudinal velocity and
rapidity of the droplets at $t_0$ are
\begin{equation}
\label{vdropz} v_{dz}=z_0/t_0,~~~~\eta_0=\frac{1}{2} \log \frac{t_0+z_0}{t_0 -z_0}\,,
\end{equation}
and the transverse velocity of the droplets may be expressed as
\cite{GBA83,WNZ09}
\begin{equation}
\label{vdropt} v_{d \perp}=a_T \bigg(\frac{\rho_0}{{\cal R}_{\perp 0}}\bigg)^{b_T} \sqrt{1-v_{dz}^2}\,,
\end{equation}
where $z_0$ and $\rho_0$ are the longitudinal and transverse
coordinates of the droplet centers at the breakup time $t_0$,  ${\cal
R}_{\perp 0}$ is the maximum transverse radius of the system at
$t_0$.  In Eq. (\ref{vdropt}), the quantities $a_T$ and $b_T$ are
the magnitude and power parameters of the transverse velocity which
will be determined by the data of particle transverse momentum
spectra.

In our model calculations, the initial droplet radius in droplet
local frame satisfies a Gaussian distribution with standard
deviation $\sigma_d=2.5$ fm, and the initial droplet centers are
assumed to distribute within a cylinder along the beam direction by
\cite{WNZ06,WNZ09}
\begin{eqnarray}
\label{dNr}
\frac{dN}{2\pi\rho_0\,d\rho_0} \propto
\left[1-\exp\,(-\rho_0^2/\Delta{\cal
R}_{\perp}^2)\right]\theta({\cal R}_{\perp}-\rho_0)\,,
\end{eqnarray}
where ${\cal R}_{\perp}$ and $\Delta{\cal R}_{\perp}$ are the
initial transverse radius and shell parameter of the granular source
\cite{WNZ06,WNZ09}.  Because of the longitudinal boost-invariant in
ultrarelativistic heavy ion collisions, we may obtain the initial
coordinate $z_0$ of the droplet by the longitudinal distribution
\begin{eqnarray}
\label{dNz}
\frac{dN}{dz_0}=\frac{dN}{d\eta_0}\frac{d\eta_0}{dz_0}\propto
1\cdot\frac{t_0}{t_0^2-z_0^2},~~~~|z_0| < \sqrt{t_0^2 -\tau_0^2}.
\end{eqnarray}

The evolution of the granular source after $t_0$ is the
superposition of all the evolutions of the individual droplets,
each of which is described by relativistic hydrodynamics.  In our
calculations, we use two kinds of equations of state.  One is the
entropy density with a cross over between the QGP and hadronic gas
\cite{EOS1} (EOS1, the parameters are taken as the same as in Ref.
\cite{WNZ09}).  Another is the S95p-PCE \cite{SHEN10} (EOS2).
In order to include the pions emitted directly at hadronization
and decayed from resonances later, we let the pions freeze-out
within a wide temperature region with the probability \cite{WNZ09}
\begin{eqnarray}
\label{Pt}
\frac{dP_f}{dT} \propto f_{\rm dir}\,e^{-(T_c-T)/\Delta T_{\rm dir}}
+(1-f_{\rm dir})e^{-(T_c-T)/\Delta T_{\rm dec}}\,,~~(T_c > T > 80~{\rm MeV})\,,
\end{eqnarray}
where $T_c$ is the transition temperature, $f_{\rm dir}$ is a
fraction parameter for the direct emission, $\Delta T_{\rm dir}$ and
$\Delta T_{\rm dec}$ describe the widths of temperature for the
direct and decayed pion emissions.  They are taken to be $T_c=170$
MeV, $f_{\rm dir}=0.85$, $T_{\rm dir}=10$ MeV, and $T_{\rm dec}=90$
MeV \cite{WNZ09}.

\includegraphics{ftmp.eps}
\vspace*{1.5cm}
\hangafter=0 \hangindent=8.1cm \noindent
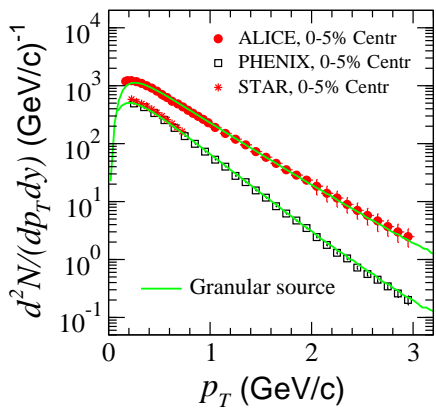
Fig. 1. (Color online) The pion transverse momentum distribution of
the granular sources and the experimental data of $\sqrt{s_{NN}}=200$
GeV Au-Au (PHENIX \cite{PHE04} and STAR \cite{STA04}) and
$\sqrt{s_{NN}}=2.76$ TeV Pb-Pb (ALICE \cite{ALI11a}) most central
collisions.
\vspace*{1.8cm}

In Fig. 1, we show the pion transverse momentum spectra
calculated with the granular source model and the experimental data
of the most central Au-Au collisions at $\sqrt{s_{NN}}=200$ GeV
\cite{PHE04,STA04} and Pb-Pb collisions at $\sqrt{s_{NN}}=2.76$ TeV
\cite{ALI11a}.  Assuming that the systems fragment when the local
energy density decreases at a certain value, we take the initial
droplet temperature $T_0=200$ MeV \cite{WNZ09} in the calculations.
For the granular source in the RHIC collisions, $\tau_0$ and the breakup
time $t_0$ are taken to be 0.8 and 4.3 fm/c.  The source parameters
${\cal R}_{\perp}$ are taken to be 6.5 fm for EOS1 and 6.8 fm for EOS2,
and $\Delta{\cal R}_{\perp}$  is taken to be 3.5 fm.  For the granular
source in the LHC collisions, these parameters are taken to be $\tau_0=
0.4$ fm/c, $t_0=8.0$ fm/c, ${\cal R}_{\perp}=$7.8(8.0) fm for EOS1(EOS2),
and $\Delta{\cal R}_{\perp}=5.4$ fm.  By comparing the transverse
momentum spectra of the granular sources with the data of $\sqrt{s_{NN}}
=200$ GeV Au-Au \cite{PHE04,STA04} and $\sqrt{s_{NN}}=2.76$ TeV Pb-Pb
\cite{ALI11a} most central collisions, we fix the parameters of the
droplet transverse velocity $a_T=$0.65(0.68) for EOS1(EOS2) for the
RHIC energy and $a_T=0.65$ for the two EOSs for the LHC energy.  The
parameter $b_T$ is fixed as 1.40 the same for the two EOSs and energies.
In Eq. (\ref{vdropt}), ${\cal R}_{\perp 0}$ is taken to be 8.0 fm in
the calculations.  It can be seen that the transverse momentum spectra
of the granular source agree well with the experimental data.

\section{HBT interferomwtry results}

In Fig. 2, we show the HBT radii $R_{\rm out}$, $R_{\rm
side}$ and $R_{\rm long}$ in the ``out", ``side", and ``long"
directions \cite{GBE88SPR90}, and the chaotic parameter $\lambda$
of the granular sources as functions of the transverse pion pair
momentum, $k_T=|{\bf p}_{1T}+{\bf p}_{2T}|/2$.  They are obtained by
fitting the two-pion correlation functions in different $k_T$
regions with the formula
\begin{equation}
C(q_{\rm out}, q_{\rm side}, q_{\rm long})=1+\lambda \,e^{-q^2_{\rm
out} R^2_{\rm out} -q^2_{\rm side} R^2_{\rm side} -q^2_{\rm long}
R^2_{\rm long}},
\end{equation}
in the longitudinal comoving system (LCMS) \cite{MAL05}.  The HBT
data of the most central collisions of $\sqrt{s_{NN}}=200$ GeV Au-Au
(STAR \cite{STA05a}) and $\sqrt{s_{NN}}=2.76$ TeV Pb-Pb (ALICE
\cite{ALI11}) are also shown.  In our calculations, we use the same
cuts as in the experimental analyses \cite{STA05a,ALI11}.  That is
the pion rapidity is limited with $|y|<0.5$ for the granular source
for the RHIC collisions and the pion pseudorapidity satisfies
$|\eta|<0.8$ for the granular source for the LHC collisions,
respectively.

One can see the HBT radii of the granular sources agree well with
the experimental data.  Because of the higher collision energy,
the granular source for the LHC collisions is formed (system
energy density decreases to a certain value) at the larger $t_0$,
and hence has a wider $z_0$ distribution (see Eq. (\ref{dNz}))
and larger ${\cal R}_{\perp}$ value (wider $\rho_0$ distribution,
see Eq. (\ref{dNr})).  These lead to the wider distributions of
the longitudinal and transverse source points for the LHC granular
source than those for the RHIC granular source \cite{WNZ11}.  The
wider distributions of the transverse and longitudinal source
points for the LHC granular source lead to the larger transverse
and longitudinal HBT radii than those for the RHIC granular source.

For a certain $k_T$ bin, the transverse velocity of the pair,
$({\bf p}_{1T}+{\bf p}_{2T})/(E_1+E_2)$, is fixed.  The ratio of
the transverse HBT radii $R_{\rm out}(k_T)/R_{\rm side}(k_T)$ is
related to the source transverse expansion, which may change the
source distribution of the correlated pions in the out and side
directions \cite{UHE02,WNZ06a}, and the average relative emitting
time of the two pions, ${\overline {\Delta t}}=\langle |t_1-t_2|
\rangle \sim [\langle(t_1-t_2)^2\rangle]^{1/2}=[\langle t_1^2
\rangle -\langle t_1 \rangle^2 + \langle t_2^2 \rangle - \langle
t_2 \rangle^2]^{1/2}$ \cite{UAW99,MAL05,SCH95}.  Although the
larger breakup time $t_0$ for the LHC source leads to a larger
particle-emitting time, the limited value of ${\overline {\Delta t}}$
for the granular sources \cite{WNZ04,WNZ06,WNZ09,WNZ06a} and the
large droplet transverse velocity lead to the small $R_{\rm out}/
R_{\rm side}$ results (see Fig. 2 (d) and (d')).  In the upper panel
of Fig. 3, we show the average pion-emitting time ${\overline
{\,t\,}}$ and the average relative pion-emitting time ${\overline
{\Delta t}}$ for the RHIC and LHC granular sources.  It can be seen
that the average pion-emitting time for the LHC source is much
larger than that for the RHIC source.  However, there is only small
difference between the average relative pion-emitting times for the
LHC and RHIC sources.  In the lower panel of Fig. 3, we show
the average source transverse and longitudinal velocities as
functions of the transverse pion pair momentum $k_T$.  The
transverse velocity ${\overline v}_{_{\rm ST}}$ increases with
$k_T$, and the longitudinal velocity ${\overline v}_{_{\rm SL}}$
decreases with $k_T$.  For the collisions at the LHC energy the
granular source has larger transverse and longitudinal velocities.

\includegraphics{fhbtrn.eps} \vspace*{8.5cm}
\hangafter=0 \hangindent=-7.8cm \noindent
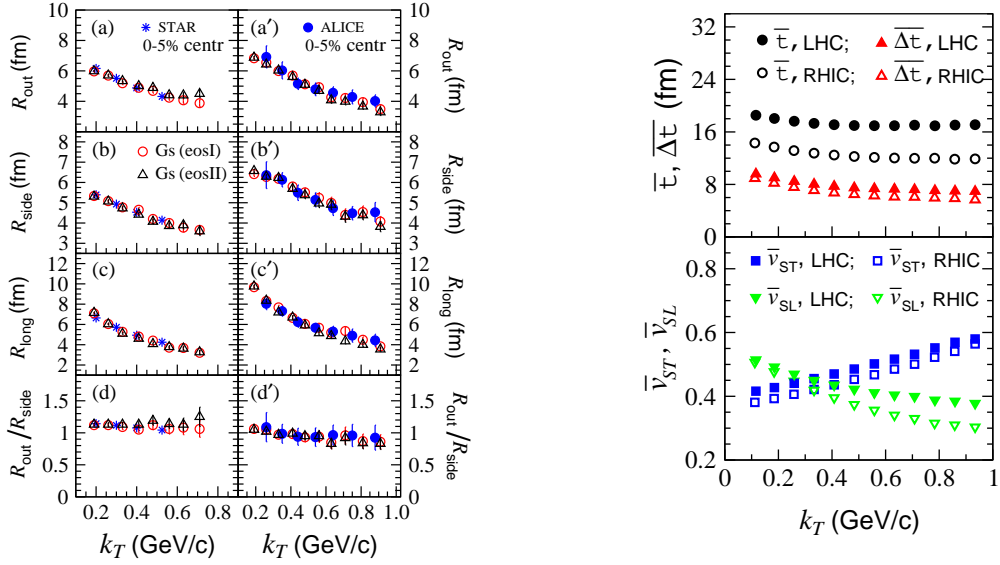
Fig. 2. (Color online) The HBT radii of the granular sources and the
experimental data for the most central collisions of $\sqrt{s_{NN}}=
200$ GeV Au-Au \cite{STA05a} and $\sqrt{s_{NN}}=2.76$ TeV Pb-Pb
\cite{ALI11}.

\vspace*{-9.6cm}
\includegraphics{ftv.eps} \vspace*{7.0cm}
\hangafter=0 \hangindent=8.5cm \noindent
Fig. 3. (Color online) (a) The average pion-emitting time and
relative pion-emitting time for the granular sources.
(b) The average source transverse and longitudinal velocities
for the granular sources.
\vspace*{-0.5cm}

\section{Summary}

In summary, we investigate the two-pion HBT interferometry in
ultrarelativistic heavy ion collisions in the granular source model
of QGP droplets.  The pion transverse momentum spectra and HBT radii
of the granular sources agree well with the data of RHIC $\sqrt{s_{NN}}
=200$ GeV Au-Au and LHC $\sqrt{s_{NN}}=2.76$ TeV Pb-Pb most central
collisions.  Our results indicate that the granular source for the
collisions at the LHC energy may have a larger initial system breakup
time as compared to the collisions at RHIC energy.  The larger
breakup time leads to the broader longitudinal and transverse
distributions of the source points, and hence leads to the larger
longitudinal and transverse HBT radii.  However, the average
relative emitting time of the two pions is limited in the granular
source model.  The limited relative pion-emitting time and larger
droplet transverse expansion lead to the ratio of the transverse
HBT radii $R_{\rm out}/R_{\rm side}\sim 1$.  In our granular
source model, there is correlations between the space-time
coordinates and source velocities.  The data of the particle
transverse momentum spectra and HBT radii for the collisions at the
RHIC and LHC energies give strong constraints for the model
parameters.  The consistent explanation of the granular source model
to these experimental data will be helpful to understand the formation
and evolution of the pion-emitting sources in ultrarelativistic
heavy ion collisions.  Further studies on the relationship of the
source radii and emission time and the comparison with other models,
which can explain the source functions of $\sqrt{s_{NN}}=200$ GeV
Au-Au \cite{PHE08}, will be interest.

\begin{acknowledgments}
This research was supported by the National Natural Science
Foundation of China under Contract No. 11075027.
\end{acknowledgments}

\end{document}